\documentclass[12pt,a4]{article}
\usepackage{amsmath}
\usepackage{amssymb}
\usepackage{epsfig}
\usepackage{subfigure}
\usepackage{hyperref}
\usepackage{mathrsfs}

\begin{document}
\pagestyle{plain}
\begin{titlepage}
\begin{center}

\vskip 1cm


{\large \bf {Two-Higgs Leptonic Minimal Flavour Violation}}

\vskip 1cm

F. J. Botella  $^a$ \footnote{fbotella@uv.es}, 
G. C. Branco  $^b$ \footnote{gustavo.branco@cern.ch and 
gbranco@ist.utl.pt}, 
M. Nebot $^a$ \footnote{Miguel.Nebot@uv.es} and M. N.
Rebelo $^c$ \footnote{margarida.rebelo@cern.ch and
rebelo@ist.utl.pt}

\end{center}

\vspace{1.0cm}

\noindent
{\it $^a$ Departament de F\' \i sica Te\`orica and IFIC,
Universitat de Val\`encia-CSIC, E-46100, Burjassot, Spain.} \\
{\it $^b$ Departamento de F\'\i sica and Centro de F\' \i sica Te\' orica
de Part\' \i culas (CFTP),
Instituto Superior T\' ecnico, Av. Rovisco Pais, P-1049-001 Lisboa,
Portugal.} \\
{\it  $^c$ Universidade T\' ecnica de Lisboa, Centro de F\' \i sica Te\' orica
de Part\' \i culas (CFTP), Instituto Superior T\' ecnico, Av. Rovisco Pais, 
P-1049-001 Lisboa, Portugal.} \\

\vskip 3cm

\begin{abstract}
We construct extensions of the Standard Model with two
Higgs doublets, where there are flavour changing 
neutral currents both in the quark and leptonic sectors,
with their strength  fixed by the fermion mixing
matrices $V_{CKM}$ and $V_{PMNS}$. These models are an
extension to the leptonic sector of the class of
models previously considered by Branco, Grimus and Lavoura, 
for the quark sector. We consider both the cases of Dirac 
and Majorana neutrinos and identify the minimal discrete
symmetry required in order to implement the models
in a natural way.
\end{abstract}

\end{titlepage}

\newpage

\section{Introduction}
Understanding the mechanism of gauge symmetry breaking is one of the
fundamental open questions in Particle Physics. Indeed, even if 
elementary scalar doublets are responsible for the breaking of the
gauge symmetry, one does not know
whether the breaking is generated by 
one, two or more scalar doublets. This question is
specially relevant in the beginning of the LHC era, with the 
prospects of experimentally probing the mechanism of gauge symmetry 
breaking.

In the Standard Model (SM) there is only one Higgs doublet and,
as a result, scalar couplings are automatically flavour diagonal
in the quark mass eigenstate basis. In multi-Higgs models, this is no longer
true and one is confronted with dangerous flavour changing 
neutral currents (FCNC) at tree level, unless one introduces a symmetry
leading to natural flavour conservation (NFC) \cite{Weinberg:1976hu},
\cite{Glashow:1976nt}, \cite{Paschos:1976ay}, in the Higgs sector
or some alternative mechanism to naturally suppress FCNC.

A possible alternative scenario for suppressing FCNC is
through the assumption that all flavour violating neutral couplings
be proportional \cite{Ant+hall}, \cite{Joshipura:1990pi}
to small entries of the Cabibbo-Kobayashi-Maskawa matrix ($V_{CKM}$)
\cite{Cabibbo:1963yz}, \cite{Kobayashi:1973fv}.
This is one of the ingredients of the
Minimal Flavour Violation (MFV) principle, introduced for the quark 
sector in Refs.~\cite{Buras:2000dm}, \cite{D'Ambrosio:2002ex}
and later on extended to the leptonic sector \cite{Cirigliano:2005ck},
\cite{Davidson:2006bd}, \cite{Branco:2006hz}. The first models of 
the MFV type, in the framework of two-Higgs doublets and 
without ad hoc assumptions, were proposed by Branco, 
Grimus and Lavoura (BGL) \cite{Branco:1996bq}. In the BGL models, 
the MFV character results from an exact discrete symmetry of the 
Lagrangian, spontaneously broken by the vacuum. Another proposal
for the structure of the scalar couplings to fermions is the
suggestion that the two Yukawa matrices are aligned in
flavour space \cite{aligned}.

Recently, we have proposed an extension of the hypothesis of MFV 
to general multi-Higgs models with special emphasis
on two Higgs doublets \cite{Botella:2009pq}.
In that work there is a detailed analysis of the conditions for the
neutral Higgs couplings to be only functions of $V_{CKM}$
elements as well as a MFV expansion for the neutral Higgs couplings
to fermions. This expansion
is built by combining the most basic elements   
\cite{Botella:2004ks} that
transform appropriately under weak basis transformations,
with terms proportional to the fermion mass matrices.

In this paper, we study how our analysis
can be extended to the leptonic sector, considering both the case where 
neutrinos are Dirac particles and the case where neutrinos are Majorana 
particles, acquiring naturally small masses through the seesaw 
mechanism. Note that this extension to the leptonic sector
is crucial in order to study stability under renormalization as
well as to do a full phenomenological analysis of BGL type models.

This paper is organized as follows: In section 2 we extend the BGL
model to the leptonic sector with the imposition of lepton number
conservation. Therefore, in this case, neutrinos are Dirac particles.
Furthermore, we show that in this case the one loop renormalization 
equations  for the corresponding Yukawa couplings, both in the 
quark and leptonic sector, obey the equations that 
guarantee the dependence of Higgs FCNC solely on functions of
the mixing matices. In this section we also deal with the question 
of the uniqueness of BGL models.  In section 3 we extend the BGL
model to the leptonic sector without the imposition of lepton number
conservation. We start by discussing the effective 
low energy scenario with Majorana neutrinos, and its stability. 
Next,  we 
analyse the leptonic sector in the seesaw framework taking into
consideration both the low and high energy couplings 
to the Higgs fields involving neutrinos. In section 4 we argue in 
favour of having the symmetry leading to MFV 
softly broken in the scalar potential. Finally, in section 
5 we present our Conclusions.

\section{Minimal Flavour Violation with Dirac Neutrinos}
\subsection{Framework}
The extension of BGL models to the leptonic sector 
depends on the neutrino character.
In this section, we analyse the leptonic sector of models that 
account for neutrino masses by enlarging the Standard Model (SM) 
through the introduction of three righthanded neutrinos  $\nu
_{R}^{0} $ while at the same time imposing total lepton number
conservation. As a result, only Dirac mass terms are generated 
and neutrinos are Dirac particles. We consider models with two
Higgs doublets such that the flavour changing neutral 
currents are controlled by the $V_{CKM}$ matrix  in the
quark sector and by the Pontecorvo-Maki-Nakagawa-Sakata (PMNS) matrix
\cite{Pontecorvo:1957qd},  \cite{Maki:1962mu}, \cite{Pontecorvo:1967fh}
in the leptonic sector.

The full Yukawa couplings are given by:
\begin{eqnarray}
\mathcal{L}_{Y} &=&-\overline{Q_{L}^{0}}\ \Gamma _{1}\Phi _{1}d_{R}^{0}-
\overline{Q_{L}^{0}}\ \Gamma _{2}\Phi _{2}d_{R}^{0}-\overline{Q_{L}^{0}}\
\Delta _{1}\tilde{\Phi }_{1}u_{R}^{0}-\overline{Q_{L}^{0}}\ \Delta _{2}
\tilde{\Phi }_{2}u_{R}^{0}  \nonumber \\
&&-\overline{L_{L}^{0}}\ \Pi _{1}\Phi _{1}l_{R}^{0}-\overline{L_{L}^{0}}\
\Pi _{2}\Phi _{2}l_{R}^{0}-\overline{L_{L}^{0}}\ \Sigma _{1}\tilde{\Phi _{1}}
\nu_{R}^{0}-\overline{L_{L}^{0}}\ \Sigma_{2}\tilde{\Phi_{2}}\nu_{R}^{0}+\text{h.c.}\ ,  \label{YukawaDirac1}
\end{eqnarray}
where  $\Gamma _{i}$,  $\Delta _{i}$ denote the Yukawa couplings
of the lefthanded quark doublets $Q_{L}^{0}$ to the righthanded
quarks $d_{R}^{0}$,  $u_{R}^{0}$  and to the Higgs doublets $\Phi_i$;  
$\Pi_i$, $\Sigma_i$ denote the couplings of the lefthanded leptonic
doublets $L^0_L$ to the righthanded charged leptons $l^0_R$, neutrinos 
$\nu^0_R$ and to the Higgs doublets. Lepton number conservation 
prevents the existence of invariant mass terms of Majorana type 
for righthanded neutrinos. These will appear
in section 3 in the seesaw framework, where the requirement of 
lepton number conservation will be relaxed.

In order to obtain a structure for  $\Gamma _{i}$,  $\Delta _{i}$ 
such that there are FCNC at tree level with strength completely
controlled by $V_{CKM}$, Branco, Grimus and Lavoura  imposed the
following symmetry on the quark and scalar sector of the Lagrangian 
\cite{Branco:1996bq}:
\begin{equation}
Q_{Lj}^{0}\rightarrow \exp {(i\alpha )}\ Q_{Lj}^{0}\ ,\qquad
u_{Rj}^{0}\rightarrow \exp {(i2\alpha )}u_{Rj}^{0}\ ,\qquad \Phi
_{2}\rightarrow \exp {(i\alpha )}\Phi _{2}\ ,  \label{S symetry up quarks}
\end{equation}
where $\alpha \neq 0, \pi$, with all other quark fields transforming 
trivially under the symmetry. The index $j$ can be fixed as either
1, 2 or 3. Alternatively the symmetry may be chosen as:
\begin{equation}
Q_{Lj}^{0}\rightarrow \exp {(i\alpha )}\ Q_{Lj}^{0}\ ,\qquad
d_{Rj}^{0}\rightarrow \exp {(i2\alpha )}d_{Rj}^{0}\ ,\qquad \Phi
_{2}\rightarrow \exp {(- i \alpha)}\Phi _{2}\ .  \label{S symetry down quarks}
\end{equation} \\
The symmetry given by Eq.~(\ref{S symetry up quarks}) leads to Higgs FCNC
in the down sector, whereas the symmetry specified by 
Eq.~(\ref{S symetry down  quarks}) leads to Higgs FCNC in the up sector.
The neutral Higgs interactions with the fermions, obtained from the quark 
sector of  Eq.~(\ref{YukawaDirac1}) are given by
\begin{eqnarray}
{\mathcal L}_Y (\mbox{neutral, quark})& = & - \overline{d_L^0} \frac{1}{v}\,
[M_d H^0 + N_d^0 R + i N_d^0 I]\, d_R^0  + \nonumber \\
&-& \overline{{u}_{L}^{0}} \frac{1}{v}\, [M_u H^0 + N_u^0 R + i N_u^0 I] \,
u_R^{0} + \text{h.c.}\ ,
\label{rep}
\end{eqnarray}
where $v \equiv \sqrt{v_1^2 + v_2^2} = (\sqrt{2} G_F)^{-1/2} \approx 
\mbox{246 GeV}$,  $G_F$ is the Fermi coupling constant 
 and $H^0$, $R$ are orthogonal combinations of the fields  $\rho_j$,
arising when one expands  \cite{Lee:1973iz} the neutral scalar fields
around their vacuum expectation values (vevs), $ \phi^0_j =  \frac{e^{i \theta_j}}{\sqrt{2}} 
(v_j + \rho_j + i \eta_j)$, choosing $H^0$ in such a way
that it has couplings to the quarks which are proportional
to the mass matrices, as can be seen from Eq.~(\ref{rep}). 
Similarly, $I$ denotes the linear combination
of $\eta_{j}$ orthogonal to the neutral Goldstone boson. 
The mass matrices $M_d$ and $M_u$ and the matrices $N_d^0$ and $N_u^0$ 
are given by:
\begin{eqnarray}
M_{d}=\frac{1}{\sqrt{2}}(v _{1}\Gamma _{1}+v _{2}e^{i\theta
}\Gamma _{2})\ ,\qquad 
M_{u}=\frac{1}{\sqrt{2}}(v _{1}\Delta
_{1}+v _{2}e^{-i\theta}\Delta _{2})\ , \\
N_d^0 = \frac{v_2 }{\sqrt{2}} \Gamma_1  - \frac{v_1 }{\sqrt{2}} 
e^{i \theta} \Gamma_2\ , \qquad
N_u^0 = \frac{v_2 }{\sqrt{2}} \Delta_1  - \frac{v_1 }
{\sqrt{2}} e^{-i \theta} 
\Delta_2\ , 
\end{eqnarray}
here  $\theta$ denotes the relative phase
of the vevs of the neutral components of $\Phi_i$.
The matrices $M_d, M_u$ are diagonalized by the usual 
bi-unitary transformations:
\begin{eqnarray}
U^\dagger_{dL} M_d U_{dR} = D_d \equiv \text{diag}\ (m_d, m_s, m_b)\ ,
\label{umu}\\
U^\dagger_{uL} M_u U_{uR} = D_u \equiv \text{diag}\ (m_u, m_c, m_t)\ .
\label{uct}
\end{eqnarray}
The flavour changing neutral currents are controlled by 
the matrices $N_d$ and $N_u$  related to  $N_d^0$ and $N_u^0$
by the following transformations:
\begin{equation}
N_d = U_{dL}^{\dagger }N_d^0U_{dR}\ , \qquad
N_u =  U_{uL}^{\dagger }N_u^0U_{uR}\ .
\end{equation}
In the  case of the symmetry given by
Eq.~(\ref{S symetry up quarks}), for $j=3$ 
there are FCNC in the down sector controlled by the matrix  
$N_d$  given by   \cite{Branco:1996bq}
\begin{equation}
(N_d)_{ij} \equiv  \frac{v_2}{v_1} (D_d)_{ij} - 
\left( \frac{v_2}{v_1} +  \frac{v_1}{v_2}\right) 
(V^\dagger_{CKM})_{i3} (V_{CKM})_{3j} (D_d)_{jj}\ . \label{24}
\end{equation}
whereas, there are no FCNC in the up sector and the coupling matrix
of the up quarks to the $R$ and $I$ fields is of the form:
\begin{equation}
N_u = - \frac{v_1}{v_2} \mbox{diag} \ (0, 0, m_t) +  \frac{v_2}{v_1}
\mbox{diag} \ (m_u, m_c, 0)\ . \label{25}
\end{equation}
In this example, the Higgs mediated FCNC are suppressed by the third 
row of the  $V_{CKM}$ matrix, therefore obeying to the additional 
constraint imposed on models designated as of the MFV type 
\cite{Buras:2010mh}.

As shown in reference \cite{Botella:2009pq}, the symmetries given by 
Eq.~(\ref{S symetry up quarks}) or by  Eq.~(\ref{S symetry down quarks})
lead to
\begin{eqnarray}
\mathcal{P}_{j}^{\gamma }\Gamma _{2} &=&\Gamma _{2}\ ,\qquad \mathcal{P}
_{j}^{\gamma }\Gamma _{1}=0\ ,  \label{g1g2} \\
\mathcal{P}_{j}^{\gamma }\Delta _{2} &=&\Delta _{2}\ ,\qquad \mathcal{P}
_{j}^{\gamma }\Delta _{1}=0\ ,  \label{d1d2}
\end{eqnarray}
where $\gamma $ stands for $u$ (up) or $d$ (down) quarks, and $\mathcal{P}
_{j}^{\gamma }$ are the projection operators defined \cite{Botella:2004ks} by
\begin{equation}
\mathcal{P}_{j}^{u}=U_{uL}P_{j}U_{uL}^{\dagger }\ ,\qquad \mathcal{P}
_{j}^{d}=U_{dL}P_{j}U_{dL}^{\dagger }\ ,  \label{projectors1}
\end{equation}
and $\left( P_{j}\right) _{lk}=\delta _{jl}\delta _{jk}$. Note that 
Eqs~(\ref{g1g2}) and (\ref{d1d2}), 
guarantee that the Higgs flavour changing neutral 
couplings can be written  in terms of quark masses and  $V_{CKM} $ 
entries \cite{Botella:2009pq}.  
This is a crucial feature for BGL models to be considered as of Minimal
Flavour Violation type (MFV). \\

In the leptonic sector, with Dirac type neutrinos, there is 
perfect analogy with the quark sector, consequently
MFV is enforced by one of the following symmetries.
Either 
\begin{equation}
L_{Lk}^{0}\rightarrow \exp {(i\alpha )}\ L_{Lk}^{0}\ ,\qquad \nu
_{Rk}^{0}\rightarrow \exp {(i2\alpha )}\nu _{Rk}^{0}\ ,\qquad \Phi
_{2}\rightarrow \exp {(i\alpha )}\Phi _{2} \ , \label{S symetry neutrinos}
\end{equation}
or
\begin{equation}
L_{Lk}^{0}\rightarrow \exp {(i\alpha )}\ L_{Lk}^{0}\ ,\qquad
l_{Rk}^{0}\rightarrow \exp {(i2\alpha )}l_{Rk}^{0}\ ,\qquad \Phi
_{2}\rightarrow \exp {(-i \alpha )}\Phi _{2} \ , \label{S symetry charged leptons}
\end{equation}
where, once again,  $\alpha \neq 0, \pi$, with all other 
leptonic fields transforming trivially under the symmetry. 
The index $k$ can be fixed as either 1, 2 or 3.

Similarly, for the leptonic sector, these symmetries imply
\begin{eqnarray}
\mathcal{P}_{k}^{\beta }\Pi _{2} &=&\Pi _{2}\ ,\qquad \mathcal{P}_{k}^{\beta
}\Pi _{1}=0 \ , \label{P1,P2} \\
\mathcal{P}_{k}^{\beta }\Sigma _{2} &=&\Sigma _{2}\ ,\qquad \mathcal{P}
_{k}^{\beta }\Sigma _{1}=0 \ , \label{s1,s2}
\end{eqnarray}
where $\beta $ stands for charged lepton or neutrino. In this case
\begin{equation}
\mathcal{P}_{k}^{l}=U_{lL}P_{k}U_{lL}^{\dagger }\ , \qquad 
\mathcal{P}_{k}^{\nu }=U_{\nu L}P_{k}U_{\nu L}^{\dagger }\ , \label{projectorsL}
\end{equation}
where $U_{\nu L}$ and $U_{lL}$ are the unitary matrices that diagonalize the
corresponding square mass matrices
\begin{eqnarray}
U_{lL}^{\dagger }M_{l}M_{l}^{\dagger }U_{lL} &=&\text{diag}\left( m_{e}^{2},m_{\mu
}^{2},m_{\tau }^{2}\right)\ ,  \nonumber \\
U_{\nu L}^{\dagger }M_{\nu }M_{\nu }^{\dagger }U_{\nu L} &=&\text{diag}\left(
m_{\nu _{1}}^{2},m_{\nu 2}^{2},m_{\nu 3}^{2}\right)\ ,
\label{massdiagonal lept}
\end{eqnarray}%
with $M_{l}$ and $M_{\nu }$ of the form
\begin{equation}
M_{l}=\frac{1}{\sqrt{2}}(v _{1}\Pi _{1}+v _{2}e^{i\theta }\Pi
_{2})\ ,\quad M_{\nu }=\frac{1}{\sqrt{2}}(v _{1}\Sigma _{1}+v
_{2}e^{-i\theta }\Sigma _{2})\ .  \label{massmatrixl1}
\end{equation}

\subsection{Renormalization Group Study}
Equations (\ref{g1g2}) and (\ref{d1d2}) together with 
Eqs.~(\ref{P1,P2}) and (\ref{s1,s2}) guarantee that the Higgs 
FCNC are functions of  fermion masses and of the CKM and PMNS matrices.
Therefore, it is crucial to guarantee 
the stability of these equations under renormalization.

The one loop renormalization group equations (RGE) for our Yukawa
couplings can be generalized from reference \cite{Ferreira:2010xe} to
\begin{eqnarray}
\mathcal{D}\Gamma _{k} &=&a_{\Gamma }\Gamma _{k}+  \nonumber \\
&&+\sum_{l=1}^{2}\left[ 3\text{Tr}\!\left( \Gamma _{k}\Gamma _{l}^{\dagger }+\Delta
_{k}^{\dagger }\Delta _{l}\right) +\text{Tr}\!\left( \Pi _{k}\Pi _{l}^{\dagger
}+\Sigma _{k}^{\dagger }\Sigma _{l}\right) \right] \Gamma _{l}+  \nonumber \\
&&+\sum_{l=1}^{2}\left( -2\Delta _{l}\Delta _{k}^{\dagger }\Gamma
_{l}+\Gamma _{k}\Gamma _{l}^{\dagger }\Gamma _{l}+\frac{1}{2}\Delta
_{l}\Delta _{l}^{\dagger }\Gamma _{k}+\frac{1}{2}\Gamma _{l}\Gamma
_{l}^{\dagger }\Gamma _{k}\right) \ ,  \label{RGE1}
\end{eqnarray}
\begin{eqnarray}
\mathcal{D}\Delta _{k} &=&a_{\Delta }\Delta _{k}+  \nonumber \\
&&+\sum_{l=1}^{2}\left[ 3\text{Tr}\!\left( \Delta _{k}\Delta _{l}^{\dagger }+\Gamma
_{k}^{\dagger }\Gamma _{l}\right) +\text{Tr}\!\left( \Sigma _{k}\Sigma _{l}^{\dagger
}+\Pi _{k}^{\dagger }\Pi _{l}\right) \right] \Delta _{l}+  \nonumber \\
&&+\sum_{l=1}^{2}\left( -2\Gamma _{l}\Gamma _{k}^{\dagger }\Delta
_{l}+\Delta _{k}\Delta _{l}^{\dagger }\Delta _{l}+\frac{1}{2}\Gamma
_{l}\Gamma _{l}^{\dagger }\Delta _{k}+\frac{1}{2}\Delta _{l}\Delta
_{l}^{\dagger }\Delta _{k}\right) \ ,  \label{RGE2}
\end{eqnarray}
\begin{eqnarray}
\mathcal{D}\Pi _{k} &=&a_{\Pi }\Pi _{k}+  \nonumber \\
&&+\sum_{l=1}^{2}\left[ 3\text{Tr}\!\left( \Gamma _{k}\Gamma _{l}^{\dagger }+\Delta
_{k}^{\dagger }\Delta _{l}\right) +\text{Tr}\!\left( \Pi _{k}\Pi _{l}^{\dagger
}+\Sigma _{k}^{\dagger }\Sigma _{l}\right) \right] \Pi _{l}+  \nonumber \\
&&+\sum_{l=1}^{2}\left( -2\Sigma _{l}\Sigma _{k}^{\dagger }\Pi _{l}+\Pi
_{k}\Pi _{l}^{\dagger }\Pi _{l}+\frac{1}{2}\Sigma _{l}\Sigma _{l}^{\dagger
}\Pi _{k}+\frac{1}{2}\Pi _{l}\Pi _{l}^{\dagger }\Pi _{k}\right) \ ,
\label{RGE3}
\end{eqnarray}
\begin{eqnarray}
\mathcal{D}\Sigma _{k} &=&a_{\Sigma }\Sigma _{k}+  \nonumber \\
&&+\sum_{l=1}^{2}\left[ 3\text{Tr}\!\left( \Delta _{k}\Delta _{l}^{\dagger }+\Gamma
_{k}^{\dagger }\Gamma _{l}\right) +\text{Tr}\!\left( \Sigma _{k}\Sigma _{l}^{\dagger
}+\Pi _{k}^{\dagger }\Pi _{l}\right) \right] \Sigma _{l}+  \nonumber \\
&&+\sum_{l=1}^{2}\left( -2\Pi _{l}\Pi _{k}^{\dagger }\Sigma _{l}+\Sigma
_{k}\Sigma _{l}^{\dagger }\Sigma _{l}+\frac{1}{2}\Pi _{l}\Pi _{l}^{\dagger
}\Sigma _{k}+\frac{1}{2}\Sigma _{l}\Sigma _{l}^{\dagger }\Sigma _{k}\right) \ ,
\label{RGE4}
\end{eqnarray}
where $\mathcal{D}\equiv 16\pi ^{2}\mu \left( d/d\mu \right) $ and $\mu $ is
the renormalization scale.  The coefficients 
$a_{\Gamma }$, $a_{\Delta }$, $a_{\Pi }$ 
and $a_{\Sigma }$ are given by \cite{Grzadkowski:1987wr}:
\begin{eqnarray} 
& a_{\Gamma }&= -8 g_s^2 -\frac{9}{4} g^2 -\frac{5}{12} {g^\prime}^2\ ,
\qquad 
a_{\Delta }=  -8 g_s^2 -\frac{9}{4} g^2 -\frac{17}{12} {g^\prime}^2\ .
\nonumber \\
&a_{\Pi }& =   -\frac{9}{4} g^2 -\frac{15}{4} {g^\prime}^2\ , \qquad
 a_{\Sigma }=  -\frac{9}{4} g^2 -\frac{3}{4} {g^\prime}^2\ ,
\end{eqnarray}
where $g_s$, $g$ and $g^\prime$ are the gauge coupling constants of
$SU(3)_c$, $SU(2)_L$ and $U(1)_Y$ respectively. 
To show that Eqs.~(\ref{g1g2}), (\ref{d1d2}),
(\ref{P1,P2}) and (\ref{s1,s2}) are stable under RGE one has to show that 
\begin{eqnarray}
\mathcal{P}_{j}^{\gamma }\left( \mathcal{D}\Gamma _{2}\right)  &=&\left( 
\mathcal{D}\Gamma _{2}\right)\ ,\qquad \mathcal{P}_{j}^{\gamma }\left( 
\mathcal{D}\Gamma _{1}\right) =0\ , \\
\mathcal{P}_{j}^{\gamma }\left( \mathcal{D}\Delta _{2}\right)  &=&\left( 
\mathcal{D}\Delta _{2}\right)\ ,\qquad \mathcal{P}_{j}^{\gamma }\left( 
\mathcal{D}\Delta _{1}\right) =0\ ,
\end{eqnarray}
\begin{eqnarray}
\mathcal{P}_{k}^{\beta }\left( \mathcal{D}\Pi _{2}\right)  &=&\left( 
\mathcal{D}\Pi _{2}\right)\ ,\qquad \mathcal{P}_{k}^{\beta }\left( \mathcal{D}%
\Pi _{1}\right) =0\ , \\
\mathcal{P}_{k}^{\beta }\left( \mathcal{D}\Sigma _{2}\right)  &=&\left( 
\mathcal{D}\Sigma _{2}\right)\ ,\qquad \mathcal{P}_{k}^{\beta }\left( 
\mathcal{D}\Sigma _{1}\right) =0\ ,
\end{eqnarray}
which guarantee that the Yukawa couplings at each 
different scale still verify  equations of the same form.  

It is interesting to notice that if one does not use the conditions 
for the leptonic sector given by Eqs.~(\ref{P1,P2}) and (\ref{s1,s2})
one is lead, for example, to:
\begin{equation}
\mathcal{P}_{j}^{\gamma }\left( \mathcal{D}\Gamma _{1}\right) =\text{Tr}\!\left( \Pi
_{1}\Pi _{2}^{\dagger }+\Sigma _{1}^{\dagger }\Sigma _{2}\right) \Gamma _{2}\ ,
\end{equation}
also, one must use the equality:
\begin{equation}
 \text{Tr}\!\left( \Pi _{1}\Pi _{2}^{\dagger
}+\Sigma _{1}^{\dagger }\Sigma _{2}\right) =0 
\label{ppss}
\end{equation}
in order to show that $\mathcal{P}_{i}^{\alpha }\left( 
\mathcal{D}\Gamma _{2}\right) =\left( \mathcal{D}\Gamma _{2}\right)$.
Clearly, Eq.~(\ref{ppss}) follows from  Eqs.~(\ref{P1,P2}) and (\ref{s1,s2}),
since in this case we have:
\begin{eqnarray}
\text{Tr}\!\left( \Pi _{1}\Pi _{2}^{\dagger }+\Sigma _{1}^{\dagger }\Sigma
_{2}\right) &=& \text{Tr}\!\left( \Pi _{1}\Pi _{2}^{\dagger }
\mathcal{P}_{j}^{\beta}+\Sigma _{1}^{\dagger }\mathcal{P}_{j}^{\beta }
\Sigma _{2}\right) = \nonumber \\
&=& \text{Tr}\!\left(
\left( \mathcal{P}_{j}^{\beta }\Pi _{1}\right) \Pi _{2}^{\dagger }+\left( 
\mathcal{P}_{j}^{\beta }\Sigma _{1}\right) ^{\dagger }\Sigma _{2}\right) =0\ ,
\end{eqnarray}
so that Eq.~(\ref{ppss}) is enforced by the MFV leptonic conditions.
It is the entire set of equations both in the quark and in the leptonic
sector that guarantee the stability of these models. 
This fact should not come as a surprise since  the relations
given by Eqs.~(\ref{g1g2}), (\ref{d1d2}), (\ref{P1,P2}) and (\ref{s1,s2})
follow from the imposition of a symmetry on the full Lagrangian.

In the quark sector there were six possible different implementations
of BGL type models. Three with FCNC in the down sector, 
each one corresponding to a different choice for the index $j$, and
three with FCNC in the up sector also for the different choices of $j$.
In the extension to the leptonic sector with Dirac neutrinos one
has another set of six different leptonic implementations obtained in
a similar fashion. In total, one may consider thirty six different MFV
models of BGL type in the case of Dirac neutrinos. \\

In Ref.~\cite{Botella:2009pq} we presented a MFV expansion for 
$N_d^0$ and $N_u^0$ built with terms proportional to 
$M_d$ and $M_u$ respectively, as well as products of terms 
which transform like $H_d$ and $H_u$ under weak basis transformations,
multiplying $M_d$ and $M_u$. We identified 
$\mathcal{P}_{j}^d$ and $\mathcal{P}_{k}^u$ as the simplest
such elements with the appropriate transformation under 
changes of weak basis. In fact,  $H_d$ and  $H_u$ can be 
decomposed as \cite{Botella:2004ks}:
\begin{equation}
H_{d(u)} = \sum_i {m^2_{d(u)}}_i P^{d(u)}_i\ .
\end{equation}

As a result we obtained, for example,
simple models of MFV type with Higgs mediated FCNC in both 
sectors, like the one given by the following equations:
\begin{eqnarray}
N^0_d =  \frac{v_2}{v_1} \ M_d -  \left( \frac{v_2}{v_1} +  \frac{v_1}{v_2}
\right) \  U_{uL}P_i U^\dagger_{uL} \ M_d\ , \\
N^0_u =   \frac{v_2}{v_1} \ M_u  -  \left( \frac{v_2}{v_1} +  \frac{v_1}{v_2}
\right) \  U_{dL}P_i U^\dagger_{dL} \ M_u \ .
\end{eqnarray}
Several different possible variations beyond BGL models 
were considered in  Ref.~\cite{Botella:2009pq}, obtained from 
different combinations of  $\mathcal{P}_{j}^d$ and $\mathcal{P}_{k}^u$ 
with  $M_d$ and $M_u$. We pointed out in  Ref.~\cite{Botella:2009pq}
that the zero texture structure of these models is
more involved than in the BGL case and that the question of assuring 
its loop stability, through the introduction of
symmetries, was not obvious. We can now address this question 
with the help of the renormalization group equations
for the Yukawa couplings given by Eqs.~ (\ref{RGE1}),
(\ref{RGE2}), (\ref{RGE3}) and (\ref{RGE4}).
All these additional MFV models, as well as the BGL models, lead to  
relations of the following form:
\begin{eqnarray}
\mathcal{P}_{i}^{\alpha }\Gamma _{2} &=&\Gamma _{2}\ ,\qquad \mathcal{P}
_{i}^{\alpha }\Gamma _{1}=0\ ,  \label{g1g2ge} \\
\mathcal{P}_{j}^{\beta }\Delta _{2} &=&\Delta _{2}\ ,\qquad \mathcal{P}
_{j}^{\beta }\Delta _{1}=0\ ,  \label{d1d2ge}
\end{eqnarray}
and similar equations for the leptonic sector. For $\alpha =\beta $ and $i=j$
we are in a BGL model in the quark sector. Models with $\alpha \neq \beta $
or  $i\neq j$ correspond to additional cases presented in 
Ref.~\cite{Botella:2009pq}. \\

It can be readily verified that, in general
\begin{equation}
\mathcal{P}_{i}^{\alpha }\left( \mathcal{D}\Gamma _{1}\right) =-\frac{3}{2}
\mathcal{P}_{i}^{\alpha }\Delta _{1}\Delta _{1}^{\dagger }\Gamma _{1}-2
\mathcal{P}_{i}^{\alpha }\mathcal{P}_{j}^{\beta }\Delta _{2}\Delta
_{1}^{\dagger }\mathcal{P}_{i}^{\alpha }\Gamma _{2}+\frac{1}{2}\mathcal{P}
_{i}^{\alpha }\mathcal{P}_{j}^{\beta }\Delta _{2}\Delta _{2}^{\dagger }
\mathcal{P}_{j}^{\beta }\Gamma _{1}\ .  \label{VarRGEga1}
\end{equation}
We have already shown that for BGL models we have
\begin{equation}
\mathcal{P}_{i}^{\alpha }\left( \mathcal{D}\Gamma _{1}\right) =0\ .
\end{equation}
In the case 
$\alpha =\beta $, $i\neq j$ we have 
\begin{equation}
\mathcal{P}_{i}^{\alpha }\mathcal{P}_{j}^{\beta }=0\ ,
\end{equation}
due to the fact that these  are projection operators. So we are left with 
\begin{equation}
\mathcal{P}_{i}^{\alpha }\left( \mathcal{D}\Gamma _{1}\right) =-\frac{3}{2}
\mathcal{P}_{i}^{\alpha }\Delta _{1}\Delta _{1}^{\dagger }\Gamma _{1}\ ,
\end{equation}
which, in general, is  different from zero. Therefore we conclude  
that this type of models cannot be enforced
by symmetries. The consideration of equation 
$\mathcal{P}_{i}^{\alpha }\left( \mathcal{D}\Gamma _{2}\right) =\left( 
\mathcal{D}\Gamma _{2}\right) $ would lead to similar difficulties 
and therefore, would allow us to draw a similar conclusion.
As a result we may conclude that out of the  models described by
Eqs.~(\ref{g1g2ge}) and (\ref{d1d2ge}) and their generalization to 
the leptonic sector, only BGL type models can be enforced by some symmetry. 
The same question was recently addressed  in Ref.~\cite{Ferreira:2010ir}
following a different approach. There it was shown that BGL models are
the only ones that survive among a large set of models enforced
by abelian symmetries.

\section{Minimal Flavour Violation with Majorana Neutrinos}

\subsection{Low Energy Effective Theory and Stability}
In the previous section, we assume that neutrinos are Dirac
particles. An alternative possibility is to allow for
lepton nonconservation leading to an effective Majorana mass term 
for the three light neutrinos of the form
\begin{equation}
{\mathcal L}_{\mbox{Majorana}} = 
\frac{1}{2}  {\nu_L^{0}}^T C^{-1} m_\nu \nu_L^{0} + \text{h.c.}\ ,
\end{equation}
which violates lepton number. Such a mass term is generated after 
spontaneous gauge symmetry breaking from an effective dimension five
operator $\mathcal{O}$ which, in the two Higgs doublet model 
can be written as:
\begin{equation}
 \mathcal{O} =  \sum_{i,j = 1}^2 \quad \sum_{\alpha, \beta = e, \mu, \tau}\
\quad \sum_{a, b, c, d = 1}^2
\left( L_{\mathrm{L} \alpha a}^T \kappa^{(ij)}_{\alpha \beta}
C^{-1} L_{\mathrm{L} \beta c} \right)
\left( \varepsilon^{ab} \phi_{ib} \right)
\left( \varepsilon^{cd} \phi_{jd} \right).
\label{Op}
\end{equation}
This operator contains two lefthanded lepton doublets and 
two Higgs doublets and can be viewed, for example, as arising from the
seesaw mechanism after integrating out the heavy degrees
of freedom. In the seesaw context the heavy degrees of
freedom are the righthanded neutrinos. The seesaw framework 
will be analysed in the next subsection. 

In this context we have, in the leptonic sector, the 
two flavour structures introduced before:
\begin{eqnarray}
\mathcal{L}_{Y_l}= 
-\overline{L_{L}^{0}}\ \Pi _{1}\Phi _{1}l_{R}^{0}-\overline{L_{L}^{0}}\
\Pi _{2}\Phi _{2}l_{R}^{0} + \text{h.c.}\ ,
\label{fg}
\end{eqnarray}
together with the four new flavour structures given by 
the $\kappa^{(ij)}$ matrices. 
A priori, it looks more difficult to implement
MFV in the case of Majorana neutrinos. However, this can be done
by imposing the following $Z_4$ symmetry in the effective Lagrangian   
including the terms given by Eqs.~(\ref{Op}) and (\ref{fg}):
\begin{equation}
L_{Lj}^{0}\rightarrow \exp {(i\alpha )}\ L_{Lj}^{0}\ ,\qquad
\Phi_{2}\rightarrow \exp {(i \alpha )}\Phi _{2}\ ,  \label{majsym}
\end{equation}
with $\alpha = \pi/2$. Imposing this $Z_4$ symmetry implies:
\begin{equation}
 \kappa^{(12)} = \kappa^{(21)} =  \left[\begin{array}{ccc}
0  & 0 & 0 \\
0 & 0 &  0 \\
0 & 0 & 0
\end{array}\right]\ ,
\end{equation}
and taking for definiteness $j=3$ we get 
\begin{equation}
 \kappa^{(11)} =  \left[\begin{array}{ccc}
\times  & \times & 0 \\
\times & \times &  0 \\
0 & 0 & 0
\end{array}\right]\ , \qquad
 \kappa^{(22)} =  \left[\begin{array}{ccc}
0  & 0 & 0 \\
0 & 0 &  0 \\
0 & 0 & \times
\end{array}\right]\ ,
\end{equation}
fixing the angle $\alpha$ as $\pi/2$ ensures
that $\kappa_{33}^{(22)} \neq 0$ so that the determinant of the
resulting neutrino mass matrix does not vanish automatically.
The Majorana mass matrix for the neutrinos is given by:
\begin{equation}
\frac{1}{2} m_\nu = \frac{1}{2} v_1^2 \kappa^{(11)} +
\frac{1}{2} v_2^2 e^{2i\theta} \kappa^{(22)}\ .
\end{equation}
This $Z_4$ symmetry also implies the following structure for
$\Pi_1$ and $\Pi_2$:
\begin{equation}
 \Pi_1 =  \left[\begin{array}{ccc}
\times  & \times & \times \\
\times & \times &  \times \\
0 & 0 & 0
\end{array}\right]\ , \qquad
 \Pi_2 =  \left[\begin{array}{ccc}
0  & 0 & 0 \\
0 & 0 &  0 \\
\times & \times & \times
\end{array}\right]\ .
\end{equation}
The neutrino mass matrix $m_\nu$ is block diagonal with each
block given by a different $\kappa$ matrix.  As a consequence, 
in the diagonalization of $m_\nu$, the matrices $\kappa^{(11)}$
and $\kappa^{(22)}$ are diagonalized separately. Therefore, any
linear combination of these two matrices will be simultaneously
diagonalized. As a result the lepton number violating Weinberg
operator \cite{Weinberg:1979sa}
of Eq.~(\ref{Op}) does not give rise to Higgs mediated
FCNC in the neutrino sector. For the charged lepton sector the
situation is similar to the one encountered in the previous section 
for the symmetry given by Eq.~(\ref{S symetry neutrinos}), leading 
to Higgs mediated FCNC in this sector.

The symmetry imposed by Eq.~(\ref{majsym})  in the effective 
low energy theory leads, for $j=3$ for instance, to the following conditions:
\begin{eqnarray}
\kappa^{(12)} = \kappa^{(21)} = 0\ , \qquad 
\kappa^{(11)}\mathcal{P}_{3}^{\nu} = 0\ , \qquad
\kappa^{(22)}\mathcal{P}_{3}^{\nu} = \kappa^{(22)}\ , \nonumber \\
\mathcal{P}_{3}^{\nu} \Pi_1 =0\ , \qquad \mathcal{P}_{3}^{\nu} \Pi_2 = \Pi_2 \ .
\end{eqnarray}
It can be easily verified, as we have done in section 2.2, and following the
RGE
presented in Ref.~\cite{Grimus:2004yh} that these equations are indeed stable
under renormalization, since they keep the same form at all scales.

\subsection{Seesaw Framework}
In this section, we analyse the leptonic sector in the seesaw 
framework \cite{seesaw1}, \cite{seesaw2}, \cite{seesaw3}, 
\cite{seesaw4}, \cite{seesaw5}.
We include one righthanded neutrino per generation and do not
impose lepton number conservation. The leptonic part of Yukawa couplings
and invariant mass terms can then be written: 
\begin{eqnarray}
{\mathcal L}_{Y +\mbox{mass}} & = & - \overline{L^0_L} \ \Pi_1 \Phi_1 l^0_R - 
\overline{L^0_L}\  
\Pi_2 \Phi_2 l^0_R - \overline{L^0_L} \ \Sigma_1 \tilde{ \Phi_1} \nu^0_R - 
\overline{L^0_L} \ \Sigma_2 \tilde{\Phi_2} \nu^0_R \  + \nonumber \\
& + & \frac{1}{2}  {\nu_R^{0}}^T C^{-1} M_R \nu_R^{0} +
\text{h.c.}\ .
\label{1e2}
\end{eqnarray}
The matrix $M_R$ stands for the
righthanded neutrino Majorana mass matrix. The leptonic mass matrices
generated after spontaneous gauge symmetry breaking are given by: 
\begin{eqnarray}
m_l = \frac{1}{\sqrt{2}} ( v_1 \Pi_1 + v_2 e^{i \theta} \Pi_2 )\ , \quad m_D = 
\frac{1}{\sqrt{2}} ( v_1 \Sigma_1 + v_2 e^{-i \theta} \Sigma_2)\ .
\label{mmmm}
\end{eqnarray}
Note that the notation has changed from the one in section 2,
we now have $m_l \equiv M_l$ and $m_D$ replaces $M_\nu$ in order
to avoid confusion with light neutrino masses in the seesaw framework. 
The leptonic mass terms obtained from Eq.~(\ref{1e2}) can be written as: 
\begin{equation} 
{\mathcal L}_ {\mbox{mass}} = - \overline{l_L^0}\, m_l\, l_R^0  + \frac{1}{2}  
({\nu}_{L}^{0T}, {(\nu_R^{0})^c}^T)\, C^{-1} {\mathcal M}^* 
\left( \begin{array}{c} {\nu}_{L}^{0} \\
{(\nu_R^{0})}^c \end{array} \right) + \text{h.c.}\ ,
\label{cal}
\end{equation} 
with
\begin{eqnarray}
{\mathcal M}= \left(\begin{array}{cc}
 0 & m_D \\
m^T_D & M_R \end{array}\right)\ . \label{calm}
\end{eqnarray}
We use the following convention:
\begin{equation} 
(\psi_L)^c \equiv C\gamma_0^T (\psi_L)^*\ .
\end{equation} 
The charged current couplings are given by: 
\begin{equation}
\mathcal{L}_W = - \frac{g}{\sqrt{2}} W^+_{\mu} \ \overline{l^0_L} \
\gamma^{\mu} \ \nu^0_{L} +\text{h.c.}\ .  \label{16}
\end{equation}
The neutral Higgs interactions with the fermions, obtained from
Eq.~(\ref{1e2}) can be written:
\begin{eqnarray}
{\mathcal L}_Y (\mbox{neutral, lepton})& = & - \overline{l_L^0} \frac{1}{v} \,
[m_l H^0 + N_l^0 R + i N_l^0 I]\, l_R^0  + \nonumber \\
&-& \overline{{\nu}_{L}^{0}} \frac{1}{v}\, [m_D H^0 + N_\nu^0 R + i N_\nu^0 I]\, 
\nu_R^{0} + \text{h.c.}\ ,
\label{neu}
\end{eqnarray}
with
\begin{eqnarray}
N_l^0 = \frac{v_2 }{\sqrt{2}} \Pi_1  - \frac{v_1 }{\sqrt{2}} 
e^{i \theta} \Pi_2\ ,  \\
N_\nu^0 = \frac{v_2 }{\sqrt{2}} \Sigma_1  - \frac{v_1 }
{\sqrt{2}} e^{-i \theta} 
\Sigma_2 \ .
\end{eqnarray}
There is a new feature in the seesaw framework 
due to the fact that in the neutrino sector the light 
neutrino masses are not obtained from the diagonalization of $m_D$. 

The $6 \times 6$ neutrino mass matrix ${\mathcal M}$ is diagonalized by the 
transformation:
\begin{equation}
V^T {\mathcal M}^* V = \mathscr D\ , \label{dgm}
\end{equation}
where ${\mathscr D} ={\rm diag} ({m_\nu}_1, {m_\nu}_2, {m_\nu}_3,
M_1, M_2, M_3)$,
with ${m_\nu}_i$ and $M_i$ denoting the masses of the physical
light and heavy Majorana neutrinos, respectively. 
It is convenient to write
the matrices  $V$ and $\mathscr D$ in the following block form:
\begin{eqnarray}
V=\left(\begin{array}{cc}
K & G \\
S & T \end{array}\right)\ , \qquad
{\mathscr D}=\left(\begin{array}{cc}
d & 0 \\
0 & D \end{array}\right)\ . \label{matd}
\end{eqnarray}
In the seesaw framework, with the scale of $M_R \gg v$ 
the matrix $K$ coincides to an excellent approximation with the unitary 
matrix $U$ that diagonalizes the effective mass matrix $m_{eff}$ for the 
light neutrinos: 
\begin{equation}
U^\dagger \   m_{eff} \  U^* =d  \qquad \text{with} \qquad 
m_{eff}\equiv - \ m_D \frac{1}{M_R} m^T_D\ .
\label{14}
\end{equation}
The matrix $G$ verifies the exact relation \cite{Branco:2001pq}:
\begin{equation}
G=m_D T^* D^{-1}\ , \label{exa}
\end{equation}
while $S$  is given to an excellent approximation by  \cite{Branco:2001pq}:
\begin{equation}
S^\dagger = -K^\dagger m_D M_R^{-1}\ , \label{sss}
\end{equation}
It is clear from Eqs.~(\ref{exa}) and (\ref{sss}) that $G$ and $S$ are of order
$m_D/M_R$, therefore  strongly suppressed. This in turn means that the
$3 \times 3$ matrices $K$ and $T$ are unitary to an excellent approximation.
The matrix $T$ is also very approximately determined by:
\begin{equation}
T^\dagger M_R T^* = D\ . \label{15}
\end{equation}
The physical fermion fields $l$, $\nu$ and $N$ 
are then related to the weak basis fields by:
\begin{equation}
l_L^0 = {U_l}_L l_L\,, \ \ l_R^0 =  {U_l}_R l_R\,, \ \ 
\nu_L^0 = U \nu_L +G N_L\,, \ \  \nu_R^0 = S^*\nu_L^c + T^* N_L^c\,.
\label{uuuu}
\end{equation}
In terms of physical fields the charged gauge current interactions become
\begin{equation}
{\mathcal L}_W = - \frac{g}{\sqrt{2}}  (\overline{l_{L}}
\gamma_{\mu} U_{\nu} {\nu}_L W^{\mu}  +
 \overline{l_{L}} \gamma_{\mu} Q {N}_L W^{\mu})+\text{h.c.}\ .
\label{lw}
\end{equation}
$ U_{\nu}$ denotes the Pontecorvo-Maki-Nakagawa-Sakata (PMNS) matrix
defined by the product $({U_l}_L^\dagger U)$.
The second term of ${\mathcal L}_W $ in Eq.~(\ref{lw}), with mixing given by
$Q \equiv ({U_l}_L^\dagger G)$  is suppressed by $G$ and 
involves the heavy neutrinos $N$ which are not relevant for low 
energy physics.

In general the couplings of  Eq.~(\ref{neu}) lead to arbitrary scalar 
FCNC at tree level. In order for these couplings to be completely controlled 
by the PMNS matrix we introduce the following 
$Z_4$ symmetry on the Lagrangian:
\begin{equation}
L^0_{L3} \rightarrow \exp{(i \alpha)}\  L^0_{L3}\ , \qquad
\nu^0_{R3} \rightarrow \exp{(i 2\alpha)} \nu^0_{R3}\ , \qquad
\Phi_2   \rightarrow \exp{(i \alpha)} \Phi_2\ , \label{bgl}
\end{equation}
with $\alpha = \pi/2 $ and all other fields transforming
trivially under $Z_4$. The most general matrices $\Pi_i$, $\Sigma_i$ and $M_R$ 
consistent with this $Z_4$ symmetry have the following structure:
\begin{eqnarray}
 \Pi_1 & = & \left[\begin{array}{ccc}  
\times  & \times & \times \\
\times & \times &  \times \\
0 & 0 & 0 
\end{array}\right]\ , \qquad
 \Pi_2   =  \left[\begin{array}{ccc}  
0 & 0 & 0  \\
0 & 0 & 0 \\
\times & \times &  \times 
\end{array}\right]\ , \label{gam}\\
 \Sigma_1  & = & \left[\begin{array}{ccc}  
\times  & \times & 0 \\
\times & \times &  0 \\
0 & 0 & 0 
\end{array}\right]\ , \qquad 
 \Sigma_2   =  \left[\begin{array}{ccc}  
0  & 0 & 0 \\
0 & 0 &  0 \\
0 & 0 & \times
\end{array}\right]\ , \qquad 
M_R   =   \left[\begin{array}{ccc}  
\times  & \times & 0 \\
\times & \times &  0 \\
0 & 0 & \times 
\end{array}\right]\ ,
\label{del}
\end{eqnarray}
where $\times$ denotes an arbitrary entry while the zeros are imposed
by the symmetry $Z_4$. Note that the choice of $Z_4$ is crucial 
in order to guarantee $M_{33}\neq 0$ and thus a non-vanishing
$\det M_R$. The same choice was required in the previous 
subsection in order to allow for a non-vanishing determinant for
the effective Majorana neutrino mass matrix.
In this weak basis the following important relations are 
verified:
\begin{eqnarray}
P_3 \Pi_2 =  \Pi_2\ ,  \qquad P_3 \Pi_1 = 0\ , \qquad \text{with} \qquad 
P_3   =  \left[\begin{array}{ccc}  
0 & 0 & 0  \\
0 & 0 & 0 \\
0 & 0 & 1 
\end{array}\right] \ ,
\end{eqnarray}
as well as
\begin{equation}
P_3 \Sigma_2 =  \Sigma_2\ ,  \qquad P_3 \Sigma_1 = 0\ .
\end{equation}
\subsection{Full Seesaw Higgs couplings}
Let us now write the neutral scalar couplings of the charged leptons
in the mass eigenstate basis:
\begin{multline}
{\mathcal L}_Y ^l(\mbox{neutral}) = - \frac{H^0}{v}\, \overline{l}\, D_l\, l  \\ 
- \frac{R}{v}\, \overline{l}\, (N_l \gamma_R + N_l^\dagger \gamma_L) )\, l
+ i\frac{I}{v}\, \overline{l}\, (N_l \gamma_R - N_l^\dagger \gamma_L) )\, l\ ,
\label{chd}
\end{multline}
where $\gamma_L = (1 - \gamma_5)/2$ , $\gamma_R = (1 + \gamma_5)/2$  
and $N_l \equiv  {U_l}_L^\dagger \ N_l^0 \ {U_l}_R $.

The fact that $U$ given by Eq.~(\ref{14}) is block diagonal 
with no mixing for the third family leads to:
\begin{equation}
(N_l)_{ij} \equiv  ({U_l}_L^\dagger \ N_l^0 \ {U_l}_R)_{ij} 
= \frac{v_2}{v_1} (D_l)_{ij} - 
\left( \frac{v_2}{v_1} +  \frac{v_1}{v_2}\right) 
(U^\dagger_{\nu})_{i3} (U_{\nu})_{3j} (D_l)_{jj}\ . \label{64}
\end{equation}
$U_{\nu}$ is the PMNS matrix.

We obtain the neutrino couplings to neutral scalars from the last term of 
Eq.~(\ref{neu}). It is useful to rewrite $N_\nu^0 $ in the form:
\begin{equation}
N_\nu^0 =   \frac{v_2}{v_1} m_D - \frac{v_2}{\sqrt 2}
\left( \frac{v_2}{v_1} +  \frac{v_1}{v_2}\right) e^{-i \theta} \Sigma_2\ .
\label{ola}
\end{equation}
Notice that the first term of $N_\nu^0$
is proportional to $m_D$ and therefore these couplings to the fields $R$ 
and $I$  have a structure similar to the $H^0$ couplings in Eq.~(\ref{neu}). 
The couplings of the neutrino mass eigenstates $\nu_i$ (light), $N_i$
(heavy) to the neutral scalars $H^0$, $R$ and $I$ are more involved
than the couplings of the charged leptons, since
they include light-light, light-heavy and heavy-heavy couplings. In the 
sequel, we shall consider each one of these terms, displaying their 
explicit form in the present model. \\

\noindent
{\bf \underline{$H^0$ couplings}} \\

\noindent
(i) light-light couplings. \\
These couplings can be written
\begin{equation}
{\mathcal L}_{\nu \nu}^{H^0} = \frac{A_{ij}}{v}\  \overline{{\nu}_{iL}} 
H^0 {\nu}_{jL}^c + \text{h.c.}\ ,
\end{equation}
where
\begin{equation}
A = U^\dagger m_D S^* = d\ .
\label{ad}
\end{equation}
These couplings among light neutrinos are flavour diagonal and are proportional
to the light neutrino masses.
From the point of view of the
effective low energy theory there are no scalar FCNC
in the neutrino sector, since, as will be shown, the term of $N_\nu^0$ in  
$\Sigma_2$ given by Eq.~(\ref{ola}), corresponding to light-light
couplings ($ U^\dagger  \Sigma_2S^* $)
will not generate nonzero off-diagonal entries. 
Its form is given explicitly, in the sequel, by Eq.~(\ref{uss}).\\

\noindent
(ii) light-heavy couplings. \\
We write these terms as
\begin{equation}
{\mathcal L}_{\nu N}^{H^0} = \frac{B_{ij}}{v} \ \overline{{\nu}_{Li}} 
H^0 N_{Lj}^c + \frac{E_{ij}}{v}\  \overline{N_{Li}} H^0 {\nu}_{Lj}^c
+ \text{h.c.}\ ,
\end{equation}
where
\begin{equation}
B = U^\dagger m_D T^*\ ,, \qquad  E = G^\dagger m_D S^* \ .
\label{bded}
\end{equation}
From  Eqs.~(\ref{14}) and  (\ref{15}) one can write
\begin{equation}
B = (i \sqrt{d}\, O^c \sqrt{D})\ ,
\label{bdo}
\end{equation}
where $O^c$ is an orthogonal complex matrix. This expression readily
follows from the Casas and Ibarra parametrization \cite{Casas:2001sr}.
The fact that $m_D$ as well as
$M_R$ are block diagonal implies that $O^c$ is also block diagonal
and can  be parametrized as:
 \begin{equation}
O^c = \left[\begin{array}{ccc}  
\cos Z & \pm \sin Z & 0 \\
- \sin Z  & \pm \cos Z & 0 \\
0 & 0 & 1
\end{array}\right]\ ,
\label{occ}
\end{equation}
with $Z$ complex. These couplings, given by the matrix $B$, are not 
suppressed by 
the mixing matrices but the fact that the heavy neutrino fields, $N$, have 
masses of order $M_R$ implies that they 
cannot be produced at low energies.  Using Eqs.~(\ref{exa}) and  (\ref{sss})
it can be readily verified that:
\begin{equation}
E = - D^{-1}  (i \sqrt{d}\, O^c \sqrt{D})^\dagger d\ ,
\label{edo}
\end{equation}
These couplings, given by the matrix $E$, are  suppressed by both matrices
 $G$ and $S$ therefore
they are much smaller than those given by $B$,
in addition, they also include a heavy neutrino.  \\

\noindent
(iii) heavy-heavy couplings. \\
One has for these couplings:
\begin{equation}
{\mathcal L}_{NN}^{H^0} = \frac{F_{ij}}{v} \  \overline{{N}_{Li}} 
H^0 {N}_{Lj}^c + \text{h.c.}\ ,
\end{equation}
where
\begin{equation}
F = G^\dagger m_D T^* \ ,
\label{fd}
\end{equation}
it can be readily verified that:
\begin{equation}
F = D^{-1} (i \sqrt{d}\, O^c \sqrt{D})^\dagger
(i \sqrt{d} O^c \sqrt{D})\ .
\label{fdo}
\end{equation}
These are couplings among heavy neutrinos and furthermore are suppressed
by the mixing matrix $G$. \\

\noindent
{\bf \underline{$R$ and $I$ couplings}} \\

\noindent
Concerning the neutral couplings to $R$ and $I$ the first term of  
$N_\nu^0 $ given by Eq.~(\ref{ola}) leads to currents with the same 
structure as those mediated by $H^0$. The second term of $N_\nu^0 $
leads to diagonal coupling matrices, due to the block structure
of $\Sigma_2$ given by Eq.~(\ref{del}) and the fact that, as a result
of the patterns given by Eq.~(\ref{del}) for the neutrino mass matrices,
the matrices $U$, $G$, $S$ and $T$ are block diagonal with no mixing 
in the third row and column. The additional couplings to $R$ and $I$ 
are derived by replacing $m_D$ by $\Sigma_2 $ in $A$, $B$, $E$ and $F$
introduced by Eqs.~(\ref{ad}), (\ref{bded}) and (\ref{fd}).
From Eq.~(\ref{ad}) and the definition of $m_D$ given by Eq.~(\ref{mmmm})
we obtain the following additional term for light-light  
couplings to $R$ and $I$:
\begin{equation}
\frac{1}{\sqrt{2}} v_2 e^{-i \theta} U^\dagger  \Sigma_2S^* = 
 \left( \begin{array}{ccc}  
0  & 0 & 0 \\
0 & 0 &  0 \\
0 & 0 & d_3
\end{array}\right)\ .
\label{uss}
\end{equation}
From Eqs.~(\ref{bded}), (\ref{bdo}) and (\ref{edo}) we obtain the following 
light-heavy coupling terms:
\begin{eqnarray}
\frac{1}{\sqrt{2}} v_2 e^{-i \theta}U^\dagger \Sigma_2 T^* = 
i  \left( \begin{array}{ccc}  
0  & 0 & 0 \\
0 & 0 &  0 \\
0 & 0 & \sqrt{d_3}\sqrt{D_3}  
\end{array}\right)\ ,  \\
\frac{1}{\sqrt{2}} v_2 e^{-i \theta} G^\dagger \Sigma_2 S^*  = 
i  \left( \begin{array}{ccc}  
0  & 0 & 0 \\
0 & 0 &  0 \\
0 & 0 & \frac{d_3}{D_3} \sqrt{d_3}\sqrt{D_3}  
\end{array}\right)\ .
\end{eqnarray}
Finally from Eqs.~(\ref{fd}), (\ref{fdo}) we obtain the following heavy-heavy coupling
term:
\begin{equation}
\frac{1}{\sqrt{2}} v_2 e^{-i \theta} G^\dagger \Sigma_2 T^* = 
 \left( \begin{array}{ccc}  
0  & 0 & 0 \\
0 & 0 &  0 \\
0 & 0 & d_3
\end{array}\right)\ .
\label{gst}
\end{equation}

Combining the two contributions from $N^0_\nu$,  the light-light couplings to 
$R$ and $I$ are diagonal. For the first two generations the coefficients 
are given 
by $\frac{v_2}{v_1} \frac{d_1}{v}$ and  $\frac{v_2}{v_1} \frac{d_2}{v}$ 
respectively. For the third generation it is given by  
$-\frac{v_1}{v_2} \frac{d_3}{v}$ .

The light-heavy couplings to $R$ and $I$ are block diagonal. Compared to the 
corresponding $H^0$ couplings the block (12) is multiplied by the ratio of vevs
$\frac{v_2}{v_1}$ and the (33) coupling is multiplied by  $-\frac{v_1}{v_2}$.
Likewise for the heavy-heavy couplings to $R$ and $I$. \\ 

From the point of view of low energy physics, the example given is a model 
of BGL type, with no Higgs mediated FCNC in the up sector (light neutrinos) 
and with the strength of the FCNC in the down sector controlled by the 
PMNS mixing matrix. All flavour changing neutral couplings with heavy neutrinos
are parametrized by both light and heavy neutrino masses,  
and the product of matrices $i \sqrt{d} O^c \sqrt{D}$, with $O^c$ of the form
given by  Eq.~(\ref{occ}). Heavy neutrino decays may be the source of the
baryon asymmetry of the universe through leptogenesis \cite{Fukugita:1986hr} 
with sphaleron processes \cite{Klinkhamer:1984di}, \cite{Kuzmin:1985mm}. \\

Next we write the Yukawa couplings to the charged Higgs, $H^+$ using 
Eqs.~(\ref{uss}) and  (\ref{gst}). \\

\noindent
{\bf \underline{$H^+$ couplings}} \\

The charged Higgs interactions with the fermions, obtained from
Eq.~(\ref{1e2}) are given by
\begin{equation}
{\mathcal L}_Y (\mbox{charged}) =  \frac{\sqrt{2} H^+}{v}
(\overline{{\nu}_{L}^{0}} N_l^0 l_R^0 +  
\overline{{\nu}_{R}^{0}}{N_\nu^0}^\dagger l_L^0) + \text{h.c.}\ .
\label{char}
\end{equation}
In the fermion  mass eigenstate basis these interactions become:
\footnotesize
\begin{eqnarray}
{\mathcal L}_Y (\mbox{charged}) & = &  \frac{\sqrt{2} H^+}{v} [
\overline{{\nu}_{L}} U_{\nu}^\dagger  N_l l_R + 
\overline{{N}_{L}} Q^\dagger  N_l l_R ] + \nonumber \\
& + &  \frac{\sqrt{2} H^+}{v} \overline{{\nu}_{L}^c}
 \left( \begin{array}{ccc}  
\frac{v_2}{v_1} d_1  & 0 & 0 \\
0 & \frac{v_2}{v_1} d_2 &  0 \\
0 & 0 & -\frac{v_1}{v_2} d_3
\end{array} \right)  U_{\nu}^\dagger l_L   +  \nonumber \\
& + & \frac{\sqrt{2} H^+}{v} \overline{{N}_{L}^c} \left(
\frac{v_2}{v_1} F^\dagger   - 
\left( \frac{v_2}{v_1} +  \frac{v_1}{v_2} \right)
 \left( \begin{array}{ccc}  
0  & 0 & 0 \\
0 & 0 &  0 \\
0 & 0 & d_3 
\end{array}\right) \right) Q^\dagger l_L \ .
\end{eqnarray}

\normalsize
\section{The scalar Potential}
The $Z_4$ symmetry which we have imposed on the Lagrangian,
forbids various gauge invariant terms in the scalar potential,
such as $ \phi_{1}^{\dagger }\phi _{2}$ , $ \phi_{1}^{\dagger }\phi _{2}
\phi_{i}^{\dagger }\phi _{i, }$, $ \phi _{1}^{\dagger }\phi _{2}
 \phi _{1}^{\dagger }\phi _{2}$. As a result, the Higgs potential has 
an exact ungauged accidental continuous symmetry, which is not a symmetry of 
the full Lagrangian. After spontaneous gauge symmetry breaking,
the accidental symmetry leads to a pseudo-Goldstone boson.
The simplest way of avoiding the pseudo-Goldstone boson
is by breaking the $Z_4$ symmetry softly through the 
introduction of the term $m_{12}   \phi_{1}^{\dagger }\phi _{2} + h.c.$.
This term avoids the pseudo-Goldstone boson which acquires a
squared mass proportional to $|m_{12}|$. In order to discuss CP
violation in this class of models, one has to consider
separately the cases of explicit and spontaneous CP violation. \\

{\bf Explicit CP violation} - If one does not impose CP invariance
at the Lagrangian level, Yukawa couplings are complex. In spite 
of the special form of these couplings, due to the presence of the 
$Z_4$ symmetry, it can be readily checked that there is in general
CP violation through the Kobayashi-Maskawa (KM) mechanism. 
The simplest way of verifying that this is the case, is by 
noting that $H_d \equiv M_d M^\dagger_d$ is a generic complex
Hermitian matrix while $H_u$ is a block diagonal matrix. One
can easily compute Tr$[H_u, H_d]^3$ and show that in general 
this weak-basis invariant does not vanish thus proving
\cite{Bernabeu:1986fc} that there is CP violation through the KM
mechanism. In order to check whether there are in this model other
sources of CP violation, one has to look at the scalar potential. 
It can be readily checked that the scalar potential, by itself, 
is CP invariant since the phase of $m_{12}$ can be removed by 
rephasing the scalar doublets, thus rendering the potential real.
The powerful Higgs-basis invariant CP-odd conditions
derived in Ref.~\cite{Branco:2005em} would obviously
provide the same answer, however this is a straightforward case.
In this variant of the model, one has all CP violation arising from the
KM mechanism. However, note that there are, for example, new
contributions to $B_d$ -- $\bar{B_d}$ apart from the usual box
diagrams of the Standard Model. These new contributions
are mediated by tree level scalar interactions, 
which are proportional to $(V_{tb}V^*_{td})^2$, therefore
with the same phase as the SM box contribution. \\

{\bf Spontaneous CP violation} -
It can be readily checked that even in the presence of the soft 
breaking term  $m_{12}   \phi_{1}^{\dagger }\phi _{2}$, one cannot 
achieve spontaneous CP violation, without enlarging the scalar 
sector. On the other hand, one may obtain spontaneous 
CP violation by introducing scalar singlets. However in order for the phase
arising from the vacuum to be able to generate a complex CKM matrix,
one has to introduce vector-like quarks \cite{vector1}.

\section{Conclusions}
We have analysed how to extend to the leptonic sector, BGL models
satisfying the minimal flavour violation (MFV) hypothesis. Both the 
cases of Dirac and Majorana neutrinos were considered. In the case
of Dirac neutrinos the extension to the leptonic is straightforward
with great similarity to the quark sector.   We have shown that
if type-I seesaw mechanism is adopted, the requirement of having
a non-singular Majorana mass matrix
for the righthanded neutrinos
further restricts the choice of the
discrete symmetry  which allows for realistic 
BGL models in the leptonic sector. 
A striking result of our analysis is the fact that 
this restricted form of the symmetry is also required
when considering the low energy effective theory with Majorana 
neutrinos. In particular, it was pointed out that BGL models 
satisfying the  MFV paradigm can be extended in a natural and elegant 
way to the leptonic sector with Majorana neutrinos, through the 
introduction of a $Z_4$ symmetry, imposed on the full Lagrangian.
Furthermore we derive the equations which guarantee calculability of
Higgs FCNC in terms of masses, $V_{CKM}$ and $V_{PMNS}$  matrices
showing that these equations are stable under renormalization. 
We have also analysed the scalar potential which acquires an exact 
ungauged accidental continuous symmetry arising from the absence of
various terms forbidden by the $Z_4$ symmetry. We have pointed 
out that the simplest way of avoiding the resulting 
pseudo-Goldstone boson is through the addition of a 
quadratic term in the scalar potential, thus softly breaking 
the $Z_4$ symmetry. Finally, we emphasize that the relevance
of BGL models stems in good part from the fact that the most
general tree-level flavour violating neutral currents are naturally
suppressed by small $V_{CKM}$ elements like the combination
($V_{td} V_{ts}^*$).

A full analysis of BGL models is beyond the scope of this 
paper and will be presented elsewhere \cite{futuro}.
It is clear that the extension of BGL models to the leptonic sector
is essential in order to make possible the above analysis and
furthermore, to allow for a consistent analysis of the 
renormalization group evolution.

\section*{Acknowledgements}
This work was partially supported by Funda\c c\~ ao para a Ci\^ encia e
a Tecnologia (FCT, Portugal) through the projects CERN/FP/109305/2009,\\ 
PTDC/FIS/098188/2008 and
CFTP-FCT Unit 777 which are partially funded through POCTI (FEDER), 
by Marie
Curie Initial Training Network "UNILHC" PITN-GA-2009-237920,
by Accion Complementaria Luso-Espanhola
PORT2008-03 and FPA-2008-04002-E/PORTU, by European FEDER, Spanish MICINN
under grant FPA2008--02878 and GVPROMETEO 2010-056. GCB and MNR are very grateful for the
hospitality of Universitat de Val\` encia during their visits. FJB and MN are
very grateful for the hospitality of CFTP/IST Lisbon during their visits. MN thanks MICINN for a \emph{Juan
de la Cierva} contract.

\end{document}